\documentclass{IEEEcsmag}

\usepackage[colorlinks,urlcolor=blue,linkcolor=blue,citecolor=blue]{hyperref}
\usepackage{upmath}
\usepackage{listings}
\usepackage{placeins}
\usepackage[normalem]{ulem}
\usepackage[binary-units]{siunitx}
\usepackage[usenames,dvipsnames]{xcolor}
\usepackage{breakurl}
\usepackage{tabularx}

\jvol{XX}
\jnum{XX}
\paper{8}
\jmonth{July/August}
\jname{IEEE Micro}
\pubyear{2022}

\lstdefinelanguage{mlir}{
    alsoletter={\%,\#,!},
    keywordsprefix={\%},
    morekeywords={\%},
    otherkeywords={f32, i32, f64, i8},
    keywords=[3]{f32, memref, i32, f64, i8, affine_map, affine_set, iter_args},
    keywordstyle=[3],
    keywords=[4]{affine},
    keywordstyle=[4],
    showstringspaces=false,
	breaklines=true,
    breakatwhitespace=true,
    morestring=[b]",
    stringstyle=,
    moredelim=[s]{\#}{<},
    moredelim=[s]{!}{\ },
    morecomment=[l]{//},
}

\begin{document}


\title{TinyIREE: An ML Execution Environment for Embedded Systems from Compilation to Deployment}

\author{Hsin-I Cindy Liu}
\affil{Google, Mountain View, CA, USA}

\author{Marius Brehler}
\affil{Fraunhofer IML, Dortmund, Germany}

\author{Mahesh Ravishankar}
\affil{Google, Seattle, WA, USA}

\author{Nicolas Vasilache}
\affil{Google, Z{\"u}rich, Switzerland}

\author{Ben Vanik}
\affil{Google, Seattle, WA, USA}

\author{Stella Laurenzo}
\affil{Google, Seattle, WA, USA}

\mark{Paper title}

\begin{abstract}
Machine learning model deployment for training and execution has been an important topic for industry and academic research in the last decade. Much of the attention has been focused on developing specific toolchains to support acceleration hardware. In this paper, we present IREE, a unified compiler and runtime stack with the explicit goal to scale down machine learning programs to the smallest footprints for mobile and edge devices, while maintaining the ability to scale up to larger deployment targets. IREE adopts a compiler-based approach and optimizes for heterogeneous hardware accelerators through the use of the MLIR compiler infrastructure which provides the means to quickly design and implement multi-level compiler intermediate representations (IR). More specifically, this paper is focused on TinyIREE, which is a set of deployment options in IREE that accommodate the limited memory and computation resources in embedded systems and bare-metal platforms, while also demonstrating IREE's intuitive workflow that generates workloads for different ISA extensions and ABIs through LLVM.
\end{abstract}

\maketitle

\chapterinitial{Machine learning} has attracted a lot of attention in the last decade, both in industry and academic research.
As new, deeper machine learning~(ML) model architectures emerge and bigger datasets become available, the demands on the hardware are also increasing.
This has led to the development of new processor architectures that accelerate ML model training and execution~(inference).
However, to fully take advantage of such acceleration hardware, the development toolchains require new deployment and compilation flows.
In addition, special considerations must be made when deploying models to mobile and edge devices, due to system resource and power constraints.

A large portion of investment in the industry has targeted use cases that scale up to the most powerful deployment targets, leaving low-power and embedded devices poorly catered to. IREE~(\textbf{I}ntermediate \textbf{R}epresentation \textbf{E}xecution \textbf{E}nvironment) was created to address this imbalance by designing a unified compiler and runtime stack with the explicit goal to scale down to the smallest footprints while retaining its ability to scale up to larger deployments (and integrate with higher-level distributed runtimes). This focus on model portability has driven IREE to effectively target bare-metal/embedded CPUs and microcontrollers. This paper presents how that portability is achieved.

Traditionally, targeting microcontrollers by public ML inference frameworks has largely been done by kernel-based, op-by-op runtimes that are hand-adapted and optimized for a small set of frequently used operators.
One approach to ease the deployment of models to microcontrollers is TensorFlow Lite for Microcontrollers~(TFLM), a framework specifically designed to support model execution on microcontrollers \cite{tflite-micro}. It uses a small runtime library~(\SI{\sim16}{\kilo\byte} for Arm Cortex-M3) that supports a subset of TensorFlow operators, with hardware-specific, high-level operator kernel implementations. The graph optimization still utilizes the TensorFlow Lite compilation flow and performs optimizations at the operator level.

Whereas TFLM is a runtime library with limited hardware-specific kernels, compiler-based approaches have also recently been applied to accelerators and embedded systems.
One of those compilers is Glow \cite{glow}, which uses a two-phase intermediate representation~(IR) to lower a neural network graph.
It uses a high-level intermediate representation to apply domain-specific optimizations.
This is followed by a lower-level IR, allowing the compiler to apply memory-related optimizations. These optimizations include instruction scheduling, static memory allocation, and copy elimination.

Another compiler is Apache TVM~\cite{tvm}, which performs both high-level graph reconfiguration and low-level operation optimization in various intermediate representations. Although TVM targets various device backends such as CPU, GPU, DSP, etc., for bare-metal microcontrollers, a MicroTVM extension is required to support the model execution scheduling and memory management, in order to make up for the lack of an operating system.

\begin{figure*}[h]
\centering
\includegraphics[width=4in]{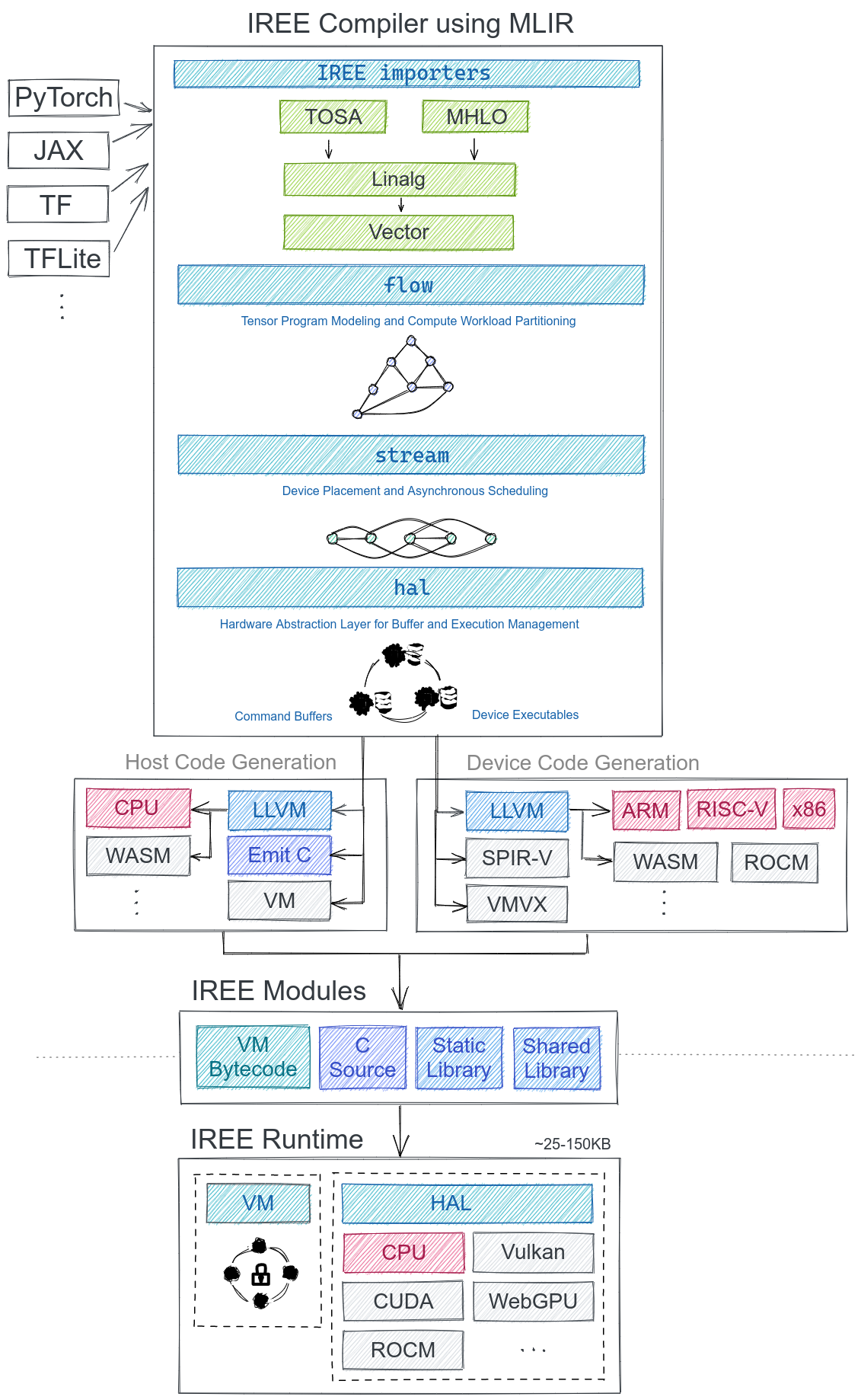}
\caption{A simplified end-to-end flow diagram of IREE.}
\label{fig:iree-overview}
\end{figure*}

This brings us to how we approach this challenge with IREE. IREE is an end-to-end compiler and runtime framework for model execution based on the MLIR compiler infrastructure~\cite{mlir}. It follows a compiler-based approach that converts ML models into an intermediate representation that allows for analysis and optimization of the ML model while generating code to target heterogeneous hardware accelerators~\cite{iree-arxiv}. As shown in \textbf{Figure \ref{fig:iree-overview}}, IREE can take various model representations as inputs, and generate executable formats for different hardware targets. IREE treats the ML model as just another program, represents it with various MLIR \emph{dialects}, and eventually generates the binary that can be run on the target architecture. An MLIR dialect can be roughly thought of as a logical level of IR. This multi-phase compilation through a progression of composable IRs~\cite{iree-arxiv} is a key differentiator because it allows us to more effectively retarget and generalize approaches and components used to target larger scale parts for specialized, deeply embedded uses. Such generalization is much harder to achieve in classic, emitter based, two phase compilers and is not possible at all in executors such as TFLM.

IREE is designed to target various edge devices and accelerators, including CPUs, GPUs, and various ML accelerators. As a result, aspects of mobile deployment, like small binary size, inter-operability with applications, and portability are first-class concerns. In this paper, we focus on TinyIREE, the subset of IREE options that generate a compact workload and runtime library that's optimized for embedded systems without an operating system. These options enable seamless support for different CPU architectures and ISA extensions, while also providing an interface to support embedded ML accelerators. In Figure~\ref{fig:iree-overview}, the TinyIREE options are colored, whereas the general IREE compilation and deployment options are grayed out.

In the following sections, we present IREE's compilation flow and explain some of the TinyIREE deployment options. Furthermore, we show results for a compiled MobileNet Single-Shot Detector~(SSD) model before giving a conclusion.

\section{IREE COMPILATION FLOW}

IREE's compilation flow uses the MLIR approach of progressively lowering an ML program into an executable for the target hardware. The primary vehicle in MLIR to achieve this is \emph{dialects}. A dialect in MLIR is a collection of operations and associated types that allow you to represent a program. Different dialects can be used to model the program at different levels of abstraction. 

For example, the IREE importers in Figure~\ref{fig:iree-overview} show some different dialects used in IREE to convert an ML program into machine executable code. The TOSA~(Tensor Operator Set Architecture)~\cite{tosa} / MHLO~(Meta High Level Optimizer) dialects model the ML program at the level of tensor operations. The Linalg dialect represents perfectly nested loop computation in a succinct, easy-to-manipulate manner, making it easy to implement transformations like fusion, tiling, loop interchange, etc. The Vector dialect models parts of the program in terms of operations on virtual vectors. Finally, the LLVM dialect is a \emph{leaf dialect} that allows MLIR programs to be translated into LLVM IR for compilation into an executable binary. Because each of these dialects are defined using a common infrastructure, at each level you can apply common compiler optimizations like Common Subexpression Elimination~(CSE), Dead Code Elimination~(DCE), etc. It is also possible to apply custom optimizations within each dialect that use the semantic information captured by operations in that dialect. For example, in the TOSA dialect, one can easily recognize \(1{\times}1\) 2D convolution operations and convert them into matrix-matrix multiply operations that can then be executed more efficiently. The process of converting from a dialect at a higher-level of representation (like TOSA/MHLO) through intermediate dialects into an executable binary---with each subsequent dialect using a more general or less specialized representation---is known as progressive lowering. The rest of the section provides an overview of the program representation for each of the dialects supported by IREE importers.

Frontend dialects like TOSA and MHLO represent the program with operations like \textit{add}, \textit{convolution}, and \textit{dot product}. For example, \textbf{Listing~\ref{lst:tosa}} shows a simple sequence with a dot product followed by bias-add in the TOSA dialect. Notice that built-in MLIR types like \texttt{tensor} use \texttt{?} to represent dynamic dimensions (i.e., dimensions whose extent is not known at compile time). As a result, MLIR has first-class support for representing dynamically shaped computations.

\begin{lstlisting}[language={mlir}, basicstyle=\ttfamily\small, basewidth={0.40em, 0.35em}, fontadjust, mathescape,caption={\small{Dot + Bias add in TOSA dialect.}},label={lst:tosa}]
%2 = "tosa.matmul"(%0, %1) : ( tensor$<$?x8xf32$>$, tensor$<$8x?xf32$>$ ) $\rightarrow$ tensor$<$?x?xf32$>$
%3 = "tosa.add"(%2, %3) : ( tensor$<$?x?xf32$>$, tensor$<$?x?xf32$>$ ) $\rightarrow$ tensor$<$?x?xf32$>$
\end{lstlisting}

The program is then lowered into the Linalg dialect. At this level, operations are represented as perfectly nested loop computations with the body of the innermost loop performing a scalar operation. This representation is intended to make it easier to apply traditional loop transformations like loop fusion, loop tiling, and loop interchange. This modeling is inspired by the polyhedral compilation domain where the computation is represented using an orthonormal space, with one axis for every loop of the computation. This orthonormal space is referred to as the \emph{iteration space} of the computation. The values of the induction variables in a loop are represented as points along the corresponding axis. Each point in this orthonormal space represents the sequence of scalar operations performed within the innermost loop. \textbf{Listing~\ref{lst:linalg}} shows the Linalg dialect's representation of a GEMM operation of the form $D_{ij} = C_{ij} + A_{ik} \cdot B_{kj}$. The number of loops is captured by the number of entries in the list \emph{iterator\_types}. The data dependence between the iterations of a loop are specified as either \emph{parallel} (for no dependence) or \emph{reduction} (for reduction-type dependence).

\begin{lstlisting}[language={mlir},basicstyle=\ttfamily\small, basewidth={0.45em, 0.35em}, fontadjust, mathescape, caption={\small{Matrix-matrix multiply in Linalg dialect.}},label={lst:linalg}]
#map0 = affine_map$<$(i, j, k) $\rightarrow$ (i, k)$>$
#map1 = affine_map$<$(i, j, k) $\rightarrow$ (k, j)$>$
#map2 = affine_map$<$(i, j, k) $\rightarrow$ (i, j)$>$
%3 = linalg.generic {
 iterator_types = ["parallel", "parallel", "reduction"],
 indexing_maps = [#map0, #map1, #map2]}
 ins(%0, %1 : tensor$<$?x?xf32$>$, tensor$<$?x?xf32$>$)
 outs(%2 : tensor$<$?x?xf32$>$) {
 ^bb0(%arg0 : f32, %arg1 : f32, %arg2 : f32):
  %0 = mulf %arg0, %arg1 : f32
  %1 = addf %arg1, %arg2 : f32
  linalg.yield %1 : f32
} $\rightarrow$ tensor$<$?x?xf32$>$
\end{lstlisting}

Similar to iteration space, each operand is also represented by an orthonormal space of the same dimensionality as the operand. This space is referred to as the \emph{data space} of the operand. For example, in a GEMM operation, all the operands are represented by a 2D data space. The maps in \emph{indexing\_maps} capture the data access pattern for each operand. In each map, the domain represents a point in the iteration space, and the range represents the point in the operand's data space that is accessed by the map.

The \emph{region} of the operation (the sequence of operations between the \{ and \}) represents the computation performed at each point in the iteration space. Each Linalg operation has a region with a single basic block whose \emph{arguments} represent the scalar values, which are obtained from the operands using the indexing map. The yielded value is the result of the computation at that point in the iteration space. For example, in Listing~\ref{lst:linalg}, each point in the iteration space performs a multiply-add operation.

All tensor operations of TOSA/MHLO can be lowered to the \texttt{linalg.generic} operation of the Linalg dialect. Two such operations that have a producer-consumer relationship can be \emph{fused} into another \texttt{linalg.generic} operation by using just the information in the iterator types and the indexing maps. This fusion encompasses all \emph{element-wise fusion} optimizations (also known as loop fusion in XLA) without having to reason about the actual computation performed by each operation. Doing the same fusion at, say, the TOSA/MHLO level, would result in a combinatorial explosion in the number of fusion possibilities.
Notice that the class of operations that can be fused in this manner is larger than pure elementwise operations, including permutations and broadcasts.

Operations in Linalg dialect can also be tiled using just the information provided by the iterator types and indexing maps. This transformation allows splitting the computation at each layer into smaller tiles of a similar computation. For example, a GEMM operation can be tiled into several smaller GEMM operations. These smaller tiles can be computed in parallel, and can therefore be distributed to different threads. Then, the computation represented by each tile is encapsulated within a \textit{dispatch region}. Each dispatch region contains code that has to be executed on the device in an atomic fashion. At this point, each computation is split into two parts for execution: The dispatch region (the tiled computation) and the code to execute host-side computation (VM commands), which specifies the order in which dispatch regions are executed. The code within dispatch regions is then transformed further using fusion, loop interchange, etc., to achieve better cache locality and data access patterns. 

Finally, the code is lowered into the \emph{Vector} dialect. This dialect represents high-level, retargetable, vector instructions available on the target hardware. Transformations within this dialect allow the generated code to use efficient vector reads/writes and special fixed function units available on the target architecture, like the Arm \texttt{sdot} instructions.

The last step in the device-side compilation is lowering the program into the LLVM dialect, which allows the program to be mechanically converted into LLVM IR. The LLVM compilation stack can then generate the binary code for the target architecture. The LLVM target flags can be used to select which CPU architecture, ABI, and ISA extensions to use. For example, to build the module for x86\_64 CPU, apply the following flag:
\begin{lstlisting}[language={bash},basicstyle=\footnotesize, basewidth={0.45em, 0.35em}, fontadjust]
- -iree-llvm-target-triple=x86_64-pc-linux-elf
\end{lstlisting}
Whereas for RISC-V 32-bit CPU with multiplication and floating point ISA extension support, use these flags:
\begin{lstlisting}[language={bash},basicstyle=\footnotesize, basewidth={0.45em, 0.35em}, fontadjust]
- -iree-llvm-target-triple=riscv32-pc-linux-elf 
- -iree-llvm-target-cpu=generic-rv32
- -iree-llvm-target-cpu-features=+m,+f
- -iree-llvm-target-abi=ilp32
\end{lstlisting}
In addition, for Armv7E-M CPU, use the following flags:
\begin{lstlisting}[language={bash},basicstyle=\footnotesize, basewidth={0.45em, 0.35em}, fontadjust]
- -iree-llvm-target-triple=armv7em-pc-linux-elf
- -iree-llvm-target-float-abi=hard
\end{lstlisting}

The aforementioned optimization process can easily be applied to other non-ML-specific linear algebra operations, as long as these operations are in the appropriate format. In other words, IREE can also optimize operations that perform input pre-processing and output post-processing as part of the program. For example, it is typical to perform color space conversion or image resizing of the image data stream for vision ML programs, and these operations can be fused within the program by following the same compilation flow.

\section{TINYIREE DEPLOYMENT OPTIONS}

The output of the compilation flow described above is an IREE module, which contains a segment of VM commands to control the buffer setup and a collection of dispatch regions as workloads. These workloads enfold the ML program that can be dispatched to the target devices. As shown in the bottom part of Figure \ref{fig:iree-overview}, when targeting embedded systems, the user needs to build an IREE runtime application, which includes the IREE runtime library to configure the VM setup, the workload loader, and a workload scheduler. The latter two are defined as the device hardware abstraction layer~(HAL) driver. 

\begin{figure}[ht!]
    \includegraphics[width=\linewidth]{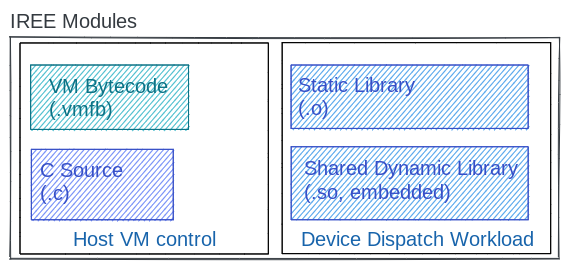}
    \caption{IREE executable deployment options for TinyIREE.}
    \label{fig:iree_target_type}
\end{figure}

Before compiling the final executable, we can further break down the IREE module structure with different deployment strategies, as described in the following subsections. As shown in \textbf{Figure~\ref{fig:iree_target_type}}, IREE provides a flexible set of tools for various deployment scenarios on CPUs, including static and dynamic embedded libraries, whereas the VM control support can be implemented with a bytecode renderer or C source code.
Here, the VM bytecode is stored in a file in FlatBuffer format (vmfb), with the option of embedding the device workload in the format of the shared dynamic library.

In addition to these deployment paths for embedded CPUs, developers can also utilize the existing HAL interface to interact with customized ML accelerators. Moreover, while IREE runtime can support dynamic deployments, embedded systems can even bypass the IREE runtime entirely and deploy a specific workload to achieve the most efficient execution.

\subsection{Virtual machine control with EmitC}
VM commands control the resource ownership and execution flow, and they are encoded as another MLIR dialect: the VM dialect.
These operations can be serialized to a bytecode module that is interpreted at runtime. Another option is to serialize these operations into C source code. This is achieved by converting the VM operations into the EmitC dialect, an MLIR dialect that allows the generation of C/C++ from MLIR. After converting the VM operations into EmitC operations, these are translated into C via the Cpp emitter. EmitC contains a call operation that represents a C function call. Amongst other things, this type of operation is used to directly call the VM API instead of serializing the calls into an equivalent bytecode representation. Therefore, the bytecode interpreter is not linked into the executable, thus reducing its size.\\
An example of using EmitC is provided in the \textit{static\_library\_demo\_c} sample in the IREE repository. 

\subsection{ML workload as a static or dynamic library}

The ML workload can be included either in a static or a dynamic library. In both cases, IREE uses LLVM to compile the program into highly optimized instruction streams for the particular target. The static ML workload library can be used in combination with the serialized VM bytecode module to form an IREE module. However, to create a binary optimized for size, the static library should instead be used in combination with the VM operations translated to C calls.

The dynamic ML workload library, on the other hand, provides more flexibility. For example, it is possible to create multiple dynamic libraries optimized for different architectures with the same entry and exit points. It is then possible to decide at runtime which of the dynamic libraries to use. Because embedded systems often do not provide dynamic library support, IREE uses a fixed dynamic library ABI, and integrates its own low-overhead dynamic library loader in the runtime library.

\subsection{Scheduler of the HAL driver}
Once the command buffer is set properly by the host VM, a workload scheduler is responsible to dispatch the device workload set in the command buffer. Inspired by the GPU scheduler and compute APIs, IREE adopts a 3D grid topology for the workload scheduling with regard to a specific worker structure that contains the data buffer and schedule information, as shown in \textbf{Listing~\ref{lst:scheduling}}. 

\begin{lstlisting}[language={C},basicstyle=\ttfamily\small, basewidth={0.45em, 0.35em}, fontadjust, mathescape,caption={Workload dispatch loop},label={lst:scheduling}]
for (int z = 0; z < worker.cnt.z; ++z) {
  for (int y = 0; y < worker.cnt.y; ++y) {
    for (int x = 0; x < worker.cnt.x; ++x) {
      vec3_t work_id = {{x, y, z}};
      int ret = dispatch_ptr(&st, &work_id);
    }
  }
}
\end{lstlisting}
For systems with multi-threading support, IREE provides an asynchronous task scheduler to split up the workload with more efficient execution of directed acyclic graphs (DAG). This allows for out-of-order and pipelined execution, which can achieve better parallelism and better utilization of the target device. However, IREE also supports synchronous scheduler execution by issuing the workload sequentially. For embedded systems without the threading handlers or a CPU without the operating system support, the synchronous scheduler provides an intuitive path to dispatch the workload.

\subsection{Stream execution}
Work scheduled by the IREE runtime is provided to the underlying system scheduler---whether a hardware command processor or CPU thread pool---with dependency information ordering the work submissions. This provides for a final level of just-in-time scheduling, work balancing, and safe preemption points for when the system is sharing limited device resources. The memory needed during the asynchronous work is reserved from a pool that is aware of the streaming behavior, and only as much memory as is required for the concurrently executable work is ever allocated. This enables deep pipelines of work to be submitted to devices without over-committing limited memory resources. Because a majority of the memory used within a typical inference is either constant or transient (local to only a particular invocation), this approach reduces the at-rest memory consumption of a loaded program to mere kilobytes, in addition to the memory required by constants, which can often be mapped into discardable memory. This allows for several programs to be loaded---even when retaining state across invocations---which can then be scheduled for interleaved execution while only consuming the peak memory used by any individual invocation.

\subsection{Buffer allocation and permission control}

The IREE runtime uses device-provided memory allocators to handle the device buffer configuration and preparation, accompanied by a system allocator to handle the host-side VM buffers. The allocators can set the visibility of the buffer between the host (VM) and the device explicitly, while using the standard memory allocation libraries for the proper alignment. A memory block can be allocated on one side (host or device), while being selectively accessible from the other end. For example, the input buffer for the ML program may be allocated at the host so it can be initiated with the correct input value, while set to be visible to the device with read access so its content can be consumed by the workload. This visibility and permission control---in conjunction with systems supporting enclave computing---provides a secured ML execution environment.

\section{RESULTS}

By applying IREE's compilation flow and choosing bare-metal-friendly configurations to deploy ML programs, we can reduce the artifact size of both the IREE module and the IREE runtime library.

\textbf{Table \ref{tab:mobilenet-workload-size}} shows the result of passing MobileNet SSD V2~\cite{mobilenetssdv2} through the IREE compilation flow for different compilation targets and modes, using IREE snapshot \href{https://github.com/google/iree/releases/tag/snapshot-20211203.686}{20211203.686}. To compile for an Armv7E-M (e.g., Arm Cortex-M4) target, the parameters are quantized. It shows both the size of the IREE module, packed into the FlatBuffer file type, and the size of the executable workload (embedded within the module FlatBuffer) across different LLVM compilation targets, each with four IREE compilation modes: Debug-dylib (dynamic library with debug symbols included; the default), Dylib (no debug symbols), Embedded (embedded system-friendly dynamic library), and Static (static library). The latter two modes are the TinyIREE modes. The result shows the workload sizes vary with the LLVM targets. In particular, the workload size is much smaller for the targets for the embedded systems, such as Armv7E-M, as a result of the LLVM optimization for such targets.

\newlength\mylength
\setlength\mylength{.35\columnwidth}
\begin{table}[h!]
    \small 
    \caption{IREE module FlatBuffer and the embedded device executable workload size for Mobilenet V2 SSD model compiled for different CPU architecture targets.}
    \begin{tabularx}{\columnwidth}{p{\mylength}rr}
    \hline
    Mode          & FlatBuffer size   & Workload size \\
                  & (\si{\kilo\byte}) & (\si{\kilo\byte}) \\
    \hline
    \textbf{x86\_64} & & \\
    Debug-dylib  & 5,588 & 256.55 \\
    Dylib        & 5,525 & 208.35 \\
    Embedded     & 5,521 & 203.80 \\
    Static       & 5,317 & -- \\
    \hline
    \textbf{Armv7E-M} & &\\
    Debug-dylib  & 5,468 & 136.55 \\
    Dylib        & 5,420 & 102.98 \\
    Embedded     & 5,406 & 88.84 \\
    Static       & 5,317 & -- \\
    \hline
    \textbf{RISC-V32imf} & & \\
    Debug-dylib  & 5,500 & 168.42 \\
    Dylib        & 5,437 & 120.18 \\
    Embedded     & 5,437 & 120.18 \\
    Static       & 5,317 & -- \\
    \hline
    \end{tabularx}
    \label{tab:mobilenet-workload-size}
\end{table}

\begin{figure}[h!]
\includegraphics[width=16pc]{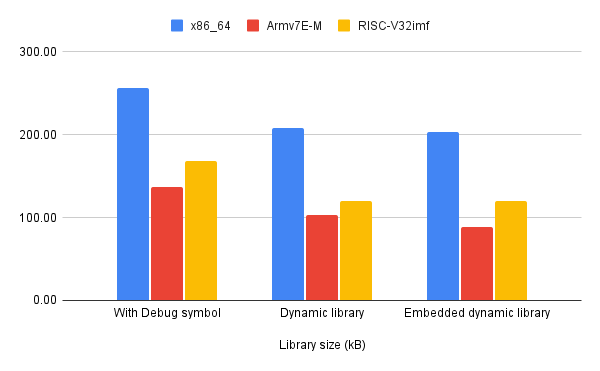}
\caption{Workload size comparison of MobileNet SSD V2 model generated by IREE flow for different CPU architectures in different deployment modes.}
\label{fig:mobilenet-library-size}
\end{figure}

\textbf{Figure \ref{fig:mobilenet-library-size}} plots the workload size comparison across different targets and modes, and the benefit of TinyIREE modes for embedded systems. Notice for the static library mode, the ML workload becomes a library that can be optimized during the link time, so its final size depends on how the user application integrates the workload library.
For example, when switching from the static library to the virtual machine control with EmitC, the bytecode interpreter can be dropped, which saves about \SI{15}{\kilo\byte} when compiling for \mbox{Armv7E-M}.

\begin{table}[h!]
    \small
    \caption{Runtime peak memory usage and the host library size for running MobileNet V2 SSD model with TFLite interpreter vs. IREE runtime application.}
    \begin{tabularx}{\columnwidth}{lrr}
    \hline
                & Peak memory (\si{\mega\byte}) & Host library (\si{\kilo\byte}) \\
    \hline
    \textbf{TFLite} &                  & \\
    x86\_64     & 20.05            & 2971.06 \\
    \hline
    \textbf{IREE}        &                  & \\
    (Embedded)  &                  & \\
    x86\_64     & 5.93            & 96.31 \\
    Armv7E-M    & 5.93            & 79.96    \\
    RISC-V32imf & 5.93            & 152.86 \\
    \hline
    \end{tabularx}
    \label{tab:mobilenet-memory}
\end{table}

In terms of execution, IREE runtime also provides more efficient memory usage, while requiring a small runtime library. \textbf{Table \ref{tab:mobilenet-memory}} shows the peak memory usage comparison between x86\_64 TensorFlow Lite~(TFLite) benchmark tool~\cite{tflite-benchmark-tool} and IREE application for Mobilenet V2 SSD model built with the dynamic embedded library mode and synchronous HAL driver. Compared with the TFLite interpreter library, the same workload can be deployed with a much smaller runtime library. Notice that TFLM could provide a host library with a similar size, but it only supports a subset of operators. Thus, not all the TFLite artifacts are compatible with TFLM. For example, the pre-2.5.0-rc0 TFLM build can deploy the same MobileNet V2 SSD model shown in Table~\ref{tab:mobilenet-memory} with~\SI{4.91}{\mega\byte} of peak memory usage and a \SI{80.02}{\kilo\byte} x86\_64 host library, but TFLM has since dropped its support for unsigned operators in favor of the recommended signed ones, hence the support for this particular model.

\section{CONCLUSION}

We presented IREE, a compiler-based ML execution framework that takes the full advantage of MLIR and that can be easily configured for different target devices. Moreover, TinyIREE options can use the same program compilation flow, but easily adjust its runtime to target embedded systems with reduced overhead. We also demonstrated the capability of supporting different CPU architectures and ISA extensions by adjusting the LLVM compilation flags. As a result, the presented framework can be optimized for multiple target backends at the same time, while keeping enough flexibility for customization.

\section{APPENDIX}

IREE is an open-source project, and we hope to encourage the ML community to contribute to the work so the framework can be applicable for more targets and applications.

Please visit \url{https://github.com/google/iree} and \url{https://github.com/iml130/iree-bare-metal-arm} to access the source code. A set of samples associated with IREE, including the quantized model shown in this paper, along with other supported ML models, are available at \url{https://github.com/google/iree-samples}.

\section{ACKNOWLEDGMENT}

We would like to thank the members of the IREE team as well as members of the team at Fraunhofer IML. Especially, we would like to thank Scott Todd for the work on runtime library development and Simon Camphausen for the work on the integration of EmitC. We would also like to thank Scott Main for the review and feedback of the manuscript.\\
The work of M. Brehler was supported by the German Federal Ministry of Education and Research (BMBF) as part of the AIA project under grant number 01IS19060A and the Competence Center Machine Learning Rhine-Ruhr (ML2R) under grant number 01IS18038B.

\begin{IEEEbiography}{Hsin-I Cindy Liu}{\,}is currently with Google, Mountain View, CA, USA. She received the Ph.~D. degree in Electrical Engineering and Computer Sciences from University of California, Berkeley in 2010. Her research interests include ambient intelligence sensing, DNN model architecture research/deployment, etc. Contact her at hcindyl@google.com.
\end{IEEEbiography}

\begin{IEEEbiography}{Marius Brehler}{\,}is currently with Fraunhofer IML, Dortmund, Germany. He received the B.Sc., M.Sc., and Dr.-Ing. degrees in Electrical Engineering from Technische Universität Dortmund, Dortmund, Germany, in 2011, 2013, and 2019, respectively. Amongst other things, his current work includes compilers for machine learning. Contact him at marius.brehler@iml.fraunhofer.de.
\end{IEEEbiography}

\begin{IEEEbiography}{Mahesh Ravishankar}{\,}is currently with Google, Seattle, WA, USA. He received his Ph.~D. degree in Computer Science and Engineering from The Ohio State University, Columbus, OH in 2014, after which he worked at NVIDIA Corporation in various teams, including the CUDA compiler team. His work primarily focuses on compilation of ML models, with a focus on code generation for CPU and GPUs. Contact him at ravishankarm@google.com.
\end{IEEEbiography}

\begin{IEEEbiography}{Nicolas Vasilache}{\,}is currently with Google, Z{\"u}rich, Switzerland. After a Ph.~D. at INRIA, France and a stint at Reservoir Labs, NYC on polyhedral compilation, he joined the Facebook AI Research Lab, NYC where he worked on fast distributed GPU libraries and led the Tensor Comprehensions systems research project. He joined Google as an early core member of MLIR in 2018 and now leads Google's Structured and Retargetable Codegen efforts in MLIR. Contact him at ntv@google.com.
\end{IEEEbiography}

\begin{IEEEbiography}{Ben Vanik}{\,}is currently with Google, Seattle, WA, USA. His experience is in real-time graphics and hardware-accelerated compute with recent work focusing on helping ML take advantage of the wealth of research that has been done in those domains. Contact him at benvanik@google.com.
\end{IEEEbiography}

\begin{IEEEbiography}{Stella Laurenzo}{\,}is currently with Google, Seattle, WA, USA. Her experience is with scaling machine learning applications and approaches to small form factors. Contact her at laurenzo@google.com.
\end{IEEEbiography}

\end{document}